\documentclass[prl,floatfix,showpacs,superscriptaddress,longbibliography,twocolumn,footinbib]{revtex4-1}

\usepackage[english]{babel}
\usepackage{amsmath,amssymb,amsfonts}
\usepackage{epsfig}
\usepackage{graphicx}
\usepackage[dvipsnames]{xcolor}

\usepackage[colorlinks=true,linkcolor=blue,citecolor=red]{hyperref}
\usepackage[normalem]{ulem}

\newcommand{\equ}[1]
{Eq.~(\ref{#1})}

\newcommand{\figu}[1]
{Fig.~\ref{#1}}

\newcount\colveccount
\newcommand*\colvec[1]{
  \global\colveccount#1
  \begin{pmatrix}
    \colvecnext
  }
  \def\colvecnext#1{
    #1
    \global\advance\colveccount-1
    \ifnum\colveccount>0
    \\
    \expandafter\colvecnext
    \else
  \end{pmatrix}
  \fi
}

\newtoks\rowvectoks
\newcommand{\rowvec}[2]{%
  \rowvectoks={#2}\count255=#1\relax
  \advance\count255 by -1
  \rowvecnexta}
\newcommand{\rowvecnexta}{%
  \ifnum\count255>0
  \expandafter\rowvecnextb
  \else
  \begin{pmatrix}\the\rowvectoks\end{pmatrix}
  \fi}
\newcommand\rowvecnextb[1]{%
  \rowvectoks=\expandafter{\the\rowvectoks&#1}%
  \advance\count255 by -1
  \rowvecnexta
}

\def\bcen{\begin{center}}
\def\ecen{\end{center}}

\def\=={\equiv}

\def\qed{\raise1pt\hbox{\vrule height5pt width5pt depth0pt}}

\def\cG0{{\cal G}_0}
\def\cG{{\cal G}}

\def\ka{{\bf k}}

 \def\=={\equiv}
 
  \def\Tr{{\rm Tr}\,}
 \def\ep0{\epsilon_{p}} \def\ed0{\epsilon_{d}}

\def\ka{{\bf k}}

\begin{document}

\title{Non-Local Annihilation of Weyl Fermions in Correlated Systems} 
\author{L.~Crippa}
\affiliation{Scuola Internazionale Superiore di Studi Avanzati (SISSA),
Via Bonomea 265, 34136
Trieste, Italy}

\author{A.~Amaricci}
\affiliation{CNR-IOM DEMOCRITOS, Istituto Officina dei Materiali,
Consiglio Nazionale delle Ricerche, Via Bonomea 265, I-34136 Trieste, Italy}
\affiliation{Scuola Internazionale Superiore di Studi Avanzati (SISSA),
Via Bonomea 265, 34136
Trieste, Italy}

\author{N.~Wagner}
\affiliation{Institut f\"ur Theoretische Physik und Astrophysik and
  W\"urzburg-Dresden Cluster of Excellence ct.qmat, Universit\"at
  W\"urzburg, 97074 W\"urzburg, Germany}

\author{G.~Sangiovanni}
\affiliation{Institut f\"ur Theoretische Physik und Astrophysik and
  W\"urzburg-Dresden Cluster of Excellence ct.qmat, Universit\"at
  W\"urzburg, 97074 W\"urzburg, Germany}

\author{J.~C.~Budich}
\affiliation{Institute of Theoretical Physics, Technische
  Universit\"at Dresden, 01062 Dresden, Germany}

\author{M. Capone}
\affiliation{Scuola Internazionale Superiore di Studi Avanzati (SISSA),
Via Bonomea 265, 34136
Trieste, Italy}

\date{ \today }

\begin{abstract}
Weyl semimetals (WSMs) are characterized by topologically stable pairs
of nodal points in the band structure, that typically originate from
splitting a degenerate Dirac point by breaking symmetries such as
time reversal or inversion symmetry. Within the independent electron approximation, the
transition between an insulating state and a WSM requires the
local creation or annihilation of one or several pairs of Weyl nodes
in reciprocal space. Here, we show that  strong electron-electron
interactions may qualitatively change this scenario. In particular, we
reveal that the transition to a Weyl semi-metallic phase can become
discontinuous, and, quite remarkably, pairs of Weyl nodes with a
finite distance in momentum space suddenly appear or disappear in the
spectral function.
We associate this behavior to the buildup of strong many-body correlations in
the topologically non-trivial regions, manifesting in dynamical
fluctuations in the orbital channel. We also highlight the impact of
electronic correlations on the Fermi arcs.
\end{abstract}


\maketitle

The classification of topological states of matter has revolutionised our understanding of
insulating~\cite{Qi2011RMP}
and nodal materials~\cite{Schoop2016NC,Young2012PRL,Young2015PRL}, 
leading to a novel conceptual framework for band-structure theory.
Regarding insulating systems, topological insulators
(TIs)~\cite{Kane2005PRL,Bernevig2006S,Fu2007PRL,Zhang2009NP,Ryu2010NJP}
represent a paradigmatic example of topological states, which are
protected by symmetries such as time reversal symmetry (TRS), and are
distinguished from conventional band insulators by topological band
structure invariants. Within the independent-electron approximation,
phase transitions between topologically distinct insulators occur via
semi-metallic Dirac points.

\begin{figure*}
  \includegraphics[width=0.9\linewidth]{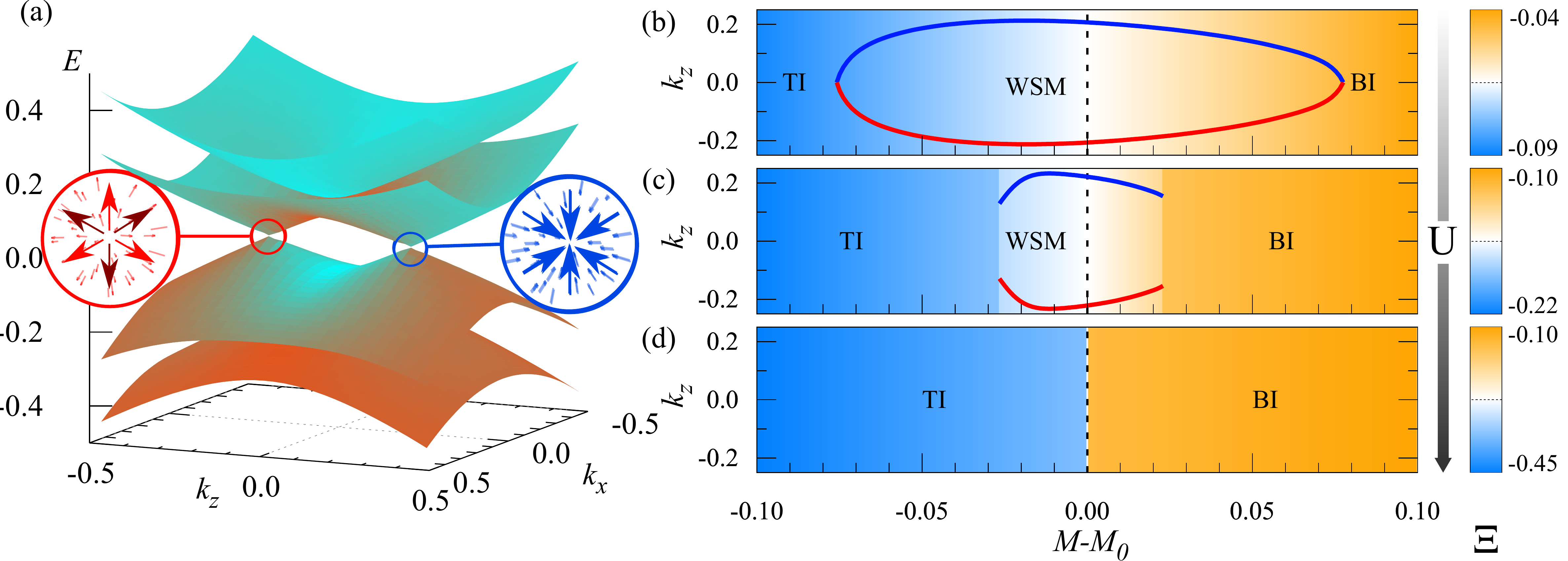}
  \caption{(Color online) 
    (a): Band-structure of  (\ref{Hmodel}) in the WSM phase 
    on $k_y\!=\!0$ plane including the $\Gamma$ point (orbital character in color). The circular insets show
     the Berry flux near the two Weyl points. 
    (b), (c), (d):
    Position of Weyl nodes along $k_z$ (for $k_x=k_y=0$ ) as a
    function of 
    $M\!-\!M_0$ for increasing values of $U$ and $b_z=0.1$. 
    $M_0$ is determined by the condition
    $M\!+\!\text{Tr}[\sigma_0\tau_z\Sigma(0)]/4=3$, where $\Sigma(0)$ is
    the self-energy function at zero frequency. This condition, which
    in the case $b_z=0$ \cite{Amaricci2015PRL} determines the topological transition point, is
    used here to center and compare data at different interactions
    strengths.   
    The background colors in the panels reflect the behavior of the correlation strength $\Xi$.  
    For small $U$ ($\approx 2.0$) the Weyl points form or annihilate continuously (b) 
    at $\Gamma$. At larger $U$ (c) they (dis)appear discontinuously
    at the boundary with an insulating state, while the width of the
    WSM phase reduces.  
    Eventually (d), for large $U$ ($\!>\!5.7$) the WSM region disappears and the
    system undergoes a direct first order
    transition between strong TI and BI.} 
%\lc{As explained in the text, effective parameters can be defined for the interacting system. At $M^{eff}=3$ a topological transition occurs in the unperturbed 3d lattice model. This value, and the correspondent bare parameter $M_{0}$ are used as a reference point to compare the different cases.} 
%  }
  \label{fig1}
\end{figure*}

In three spatial dimensions (3D), such critical Dirac points may be
promoted to extended Weyl semimetal (WSM)
phases~\cite{Shuichi_Murakami_2007,Wan2011PRB,Armitage2018RMP}
upon breaking TRS or inversion symmetry, since isolated nodal points
in the band structure are topologically stable in 3D. Concretely, a
symmetry breaking perturbation splits conventional spin-degenerate
Dirac cones into pairs of non-degenerate Weyl fermions with opposite
chirality (see Fig.~\ref{fig1}a for an illustration). At least
within the independent electron approximation such Weyl nodes can only
be {\textit{locally}} annihilated by continuously bringing them
together in reciprocal space with their chiral
antidotes~\cite{Shuichi_Murakami_2007}, and the
robustness of WSMs against interactions as well as various intriguing
correlation effects have recently been reported
\cite{AbaninPRL2014,LaubachPRB2016,Roy2017PRB,BergholtzPRB2018,MengPRL2019,PhysRevB.94.155136,Lai93,PhysRevB.98.245114,MorimotoWeylMott,VcaWSM}. WSMs 
exhibit fascinating transport phenomena, related to the chiral anomaly
and relativistic electronic
dispersion at low energy, and have recently become a major focus of theoretical
and experimental research, in a variety of materials ranging from pyrochlore iridates 
and tantalum monopnictides to synthetic materials such as optical lattices with laser-assisted tunneling \cite{Lu622,Lu2013,Yang2015,Xu2015S,
Nakatsuji2015N,Deng2016NP,Lv2015NP,Ikhlas2017NP,Jiang2017NC,Kuroda2017NMa,Lv2015PRX,PhysRevLett.122.116402,HuangTaAs,Ohtsuki2019POTNAOS}. 

In this work, we report on the discovery of drastic changes to the phenomenology of
WSMs in the presence of strong electronic
correlations (see Fig.~\ref{fig1}b for a visualisation). We show this by solving a multi-orbital
3D Hubbard model with a TRS breaking term in the framework of dynamical mean-field theory
(DMFT)~\cite{Georges1996RMP}. Quite remarkably, we find that interactions may lead to the {\textit{discontinuous}} annihilation of Weyl nodes in momentum space at a first-order WSM-to-insulator transition. In this scenario, pairs of Weyl points suddenly disappear notwithstanding their finite
separation in momentum space, leaving behind a gapped state. This genuine many-body effect may have a deep impact on the prediction and possible detection of WSMs
in correlated materials. Conversely, when approaching the transition to a WSM from the trivial band insulator (BI) phase, the nodal phase is entered discontinuously in the strongly interacting
regime, without the continuous bifurcation of a degenerate Dirac cone into two distinct Weyl nodes. Furthermore, we unveil the presence of a quantum tri-critical
point (QTP), associated to an instability in the orbital channel. 
Beyond this point the WSM region is dramatically reduced and
eventually suppressed in favor of a discontinuous transition between
two insulators.

% we find drastic
% interaction effects, that, even qualitatively, cannot be understood
% within the aforementioned single particle picture or even a
% perturbative approach to the interacting system
% \cite{BergholtzPRB2018}. 

We consider a 3D simple cubic lattice model of spinful electrons with two orbitals per site and unit lattice constant, describing a 3D TI exposed to a TRS-breaking perturbation. The Hamiltonian $H$ may be written as the sum of a single-particle term diagonal in momentum space and a
two-body interaction local in real space 
$H=\sum_\ka\Psi_\ka^\dag H_\ka\Psi_\ka +
\sum_iH^{\mathrm{int}}_i$, where we introduced the four-component spinor $\Psi(\mathbf{k})=(
\mathrm{c}_{\ka 1\uparrow}\,,
\mathrm{c}_{\ka 2\uparrow}\,,
\mathrm{c}_{\ka 1\downarrow}\,,
\mathrm{c}_{\ka 2\downarrow})$,
with the operator 
$c_{\ka\alpha\sigma}$, $\alpha=1,2$, $\sigma=\uparrow,\downarrow$
annihilating an electron with momentum $\ka$ and spin
$\sigma$ in orbital $\alpha$. The non-interacting term is
\begin{equation}
  \begin{split}
    H_\ka=&~\mathcal M(\ka)\sigma_0\tau_z+b_z\sigma_z\tau_z+\\
    &\lambda(\sin(k_x)\sigma_z\tau_x-\sin(k_y)\sigma_0\tau_y +
    \sin(k_z)\sigma_x\tau_x), 
  \end{split}
  \label{Hmodel}
\end{equation}
where $\sigma_i$ and $\tau_j$ with $i,j=x,y,z$ denote 
Pauli matrices in spin and orbital space, respectively, and $\sigma_i\tau_j$ are  Kronecker products.  
The dispersion is $\mathcal M(\ka)=M-\epsilon(\cos k_x +\cos k_y +\cos
k_z )$ and 
$b_z$ is the strength of the TRS breaking term. In the following we will consider $\epsilon$ as our unit of energy and fix $\lambda$ to $0.5$.

In the non-interacting regime and for $b_z=0$ 
the model describes a weak TI (WTI) and a strong TI
(STI) for $M<1$ and $1<M<3$, respectively, and a trivial BI for $M>3$. The
topological transitions between these phases 
occur via the continuous closure of  the energy gap and the
concomitant formation of a degenerate Dirac
cone~\cite{Amaricci2015PRL,Amaricci2016PRB}.
A finite value of  $b_{z}$ breaks the TRS and lifts the spin
degeneracy, though without giving rise to a net magnetization. 
The Dirac cone at the transition
point splits into two Weyl cones separated in momentum space along $k_z$. 
Each Weyl node acts as the magnetic (anti)monopole for the Berry
phase and is characterized by a Chern invariant measuring
its chirality~\cite{Qi2011RMP,OrderParameter,Wang2012PRX}. 
The absence of a protecting TRS makes the topological character of the WSM in
some sense more robust than the usual symmetry-protected TIs. 
In the absence of interaction the destruction (formation) of the
Weyl cones can only occur through their continuous annihilation
(separation) at a specific high-symmetry point.

The topological quantum phase transition points are thus replaced by two
distinct WSM phases in two windows $\Delta M=2b_z$  centred around 
 $M \! = \! 1$ and $M \! =\! 3$. 
The one separating the WTI from the STI features three distinct
couples of Weyl points, located near the X, X$'$ and the $\Gamma$
points and separated along the $k_z$ direction in the reciprocal lattice space.
The second WSM state, separating the STI from the
BI, features only one pair of Weyl nodes, located along $k_z$ around
$\Gamma$,  see \figu{fig1}a.  

In the Wannier basis, the Coulomb interaction is described by a set of
parameters as the on-site intra-orbital and inter-orbital terms
($U$ and $U'$) and the Hund term $J$. The couplings are related by $U' = U-2J$ for electrons
with opposite spin in the two orbitals and  $U'' = U-3J$ for electrons with the same spin.
The interaction term reads: 
\begin{equation}
\begin{split}
H_{int}=&U\sum_{il}n_{il\uparrow}n_{il\downarrow}+\\
&(U-2J)\sum_{il\neq l^{'}}n_{il\uparrow}n_{il^{'}\downarrow}+\\
&(U-3J)\sum_{il\neq l^{'}}(n_{il\uparrow}n_{il^{'}\uparrow}+n_{il\downarrow}n_{il^{'}\downarrow})
\end{split}
\label{Hint}
\end{equation}

We found that the features of the topological phase transitions are 
robust across a wide range of variations of the interaction
parameters.
Therefore, to keep the discussion well defined, in the
rest of this work we will focus on the case $J=0$, $U^{'}=U$
case. In the supplemental material \footnote{See Supplemental Material at [URL will be inserted by publisher] for a description of the effects of the parameters on the transition and the Self-Energy.} we present additional results with
the effects of the Hund's coupling $J$ and of the $U^{'}/U$ ratio on
the topological phase transition.

%To study the effect of interactions, we include a Hubbard term, neglecting the exchange term for simplicity
%$H^{\mathrm{int}}_i = \tfrac{U}{2} N_i^2$
%where $N_i = \sum_{\alpha\sigma} n_{i\alpha\sigma}$ is the total
%occupation number operator on lattice site $i$.  
The interacting model is solved non-perturbatively by means of
DMFT~\cite{Georges1996RMP,Amaricci2015PRL,Amaricci2016PRB}, i.e. by mapping the lattice problem
onto a single-site quantum impurity
coupled to an effective bath which is self-consistently determined.
Within  the DMFT approach, the effects of interaction are contained in the 
local self-energy function $\Sigma$, which in our case is a $4\times4$ matrix in the spin and orbital space retaining the
local symmetries of the problem. We obtain zero-temperature results 
using a Lanczos-based exact diagonalization
method~\cite{Georges1996RMP,Capone2007PRB,Weber2012PRB} 
discretising the bath with $N_b=9$ levels. 
The main result of this work, namely the non-local annihilation of the
Weyl points in presence of electronic interaction, is robust against a
change of $N_b$ as well as for finite temperatures up to a scale of
the order of one tenth of the energy unit. We checked the stability of
our findings comparing with a continuous-time quantum Monte Carlo
solver~\cite{w2dynamics}.

%as well as the presence of finite-$T$, \lc{up to temperature scales of the order of 1/10 of the energy unit. This has been checked both within the ED framework and with a continuous-time quantum Monte Carlo solver~\cite{w2dynamics}. It should however be remarked that the topological phase transition, being quantum in nature, are only well defined at zero temperature}.

The two aforementioned WSM phases are found to be influenced by the local Coulomb repulsion in a qualitatively distinct way. 
The one occurring between the weak- and strong TI undergoes a rather
predictable and basically adiabatic evolution as a function
of $U$. By contrast, the WSM separating the STI and the BI
drastically changes its nature as the electron-electron interaction
exceeds a certain value.  
This main result of our work is illustrated in \figu{fig1}b, c and d. The position of the Weyl points in $k_z$ as a function of $M$ is shown
by the red and blue lines. In summary, for small $U$ (\figu{fig1}b) the two nodes form a closed path, bridging the STI with
the BI region across the vertical dotted line at which, in absence of
$b_{z}$, the topological phase transition occurs. This continuous behavior of the Weyl nodes 
is adiabatically connected with the non-interacting case. 

 Surprisingly, for intermediate values of the interaction strength
(\figu{fig1}c) a discontinuous evolution is observed. 
The Weyl nodes appear all of a sudden already separated in
$k$-space at a critical value of the mass term $M$. Upon further increasing $M$,
the position of the cones evolves continuously, but they soon
non-locally disappear in favor of the band insulator, without meeting in momentum space.
Correspondingly, the red and blue lines are no longer connected, 
a situation that is fundamentally forbidden within the independent electron approximation.

\begin{figure}[t]
	\begin{center}
	\includegraphics[width=1\linewidth]{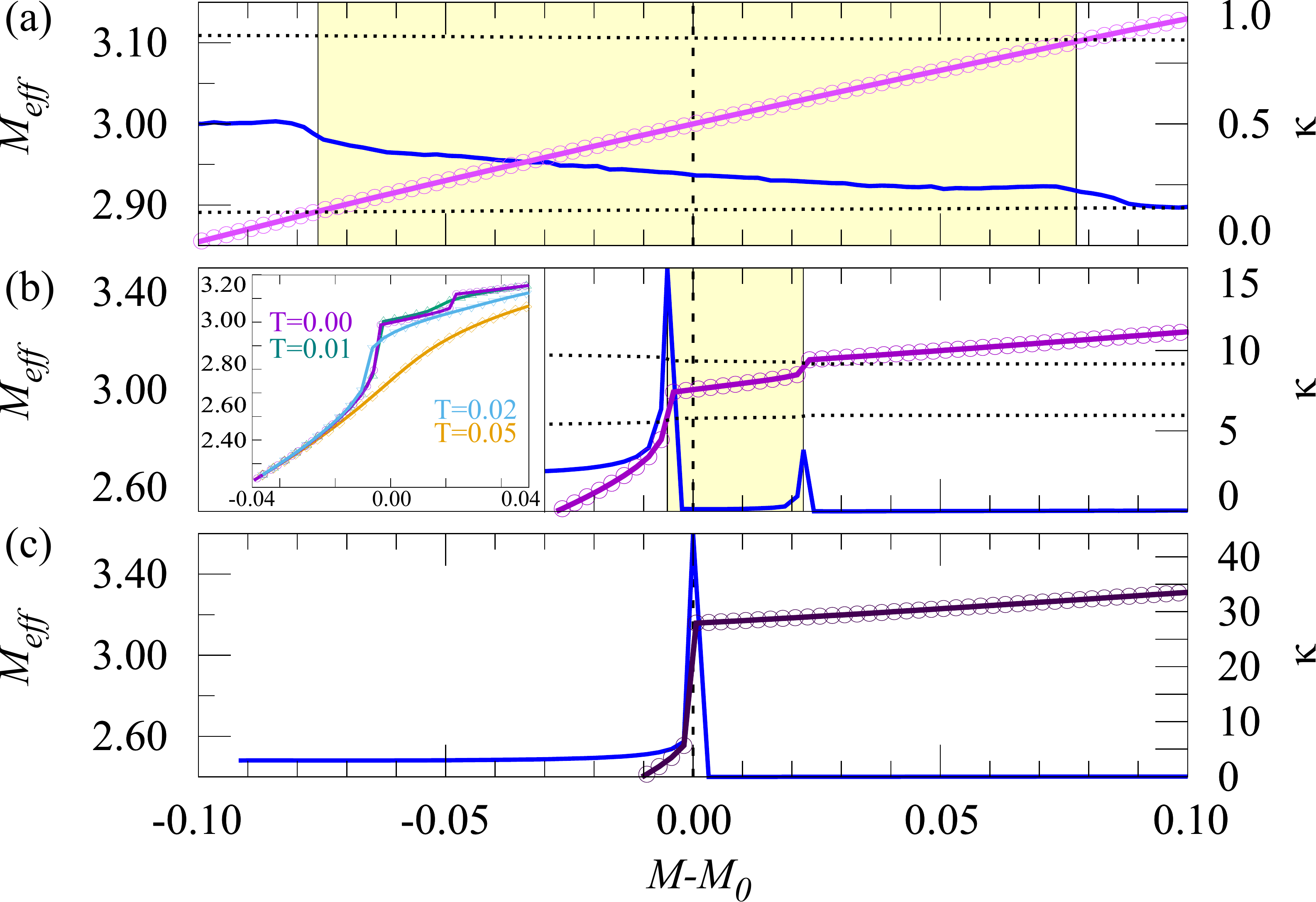}
	\caption{(Color online) $M^\text{eff}$ behaviour as a function of the bare
          mass for different values of interaction: (a) U=3.0, (b) U=5.5, (c) U=5.9. $M_0$ is defined
          as in Fig.~\ref{fig1}.
          The shaded regions
          correspond to the WSM, that of the
          noninteracting system being the  full width of the panels.  
          The dotted lines represent (where significant) the expected width of the WSM region,
          spanning the interval
          $M^\text{eff}=[3-b^\text{eff}_z,3+b^\text{eff}_z]$. The blue
          lines represent the orbital compressibility $\kappa$.
          The inset to panel (b) shows the behavior of $M^\text{eff}$ for $U\!=\!5.5$ as a function of temperature.
          }
	\label{fig2}
	\end{center}
\end{figure}

This intriguing phenomenon can be further characterized by inspecting the
correlation strength $\Xi$, which we measure in terms of the deviation from Hartree-Fock theory of 
our semi-metallic solutions~\cite{Amaricci2017PRB,Amaricci2015PRL,Amaricci2016PRB}.
Since within DMFT we take into account contributions to the (local)
self-energy to all orders, our solution displays self-energy corrections that are much
richer than Hartree-Fock theory. 
In particular, the self-energy $\Sigma$ is found to exhibit a pronounced frequency structure
at intermediate-to-large $U$,  thus qualitatively deviating from the static mean field solution \cite{Budich2012PRB,Budich2013PRB}. 
This is a consequence of the pronounced many-body fluctuations of the STI and of the correlated Weyl phases, originating from their 
orbitally entangled nature. 
%\gs{Not sure that it was clever of me mentioning the STI here....also: we are using a definition of $\Xi$ with negative values. This makes it strange to follow the evolution from the TI to the BI, as we go from something negative to something negative but smaller....}
%In addition, the non-zero value of the imaginary part of the self-energies
%away from the Fermi level in DMFT describes the finite lifetime effects of the
%excitations, contrary to the infinitely long-lived Hartree-Fock quasi-particles. 
We can quantify this by computing the difference between the low-frequency limit, 
which describes the quasiparticles, and the high-energy limit
$\Xi={\Tr{\left[\sigma_0\tau_z\Sigma(0)-
      \sigma_0\tau_z\Sigma^\mathrm{HF}\right]}}/{\Tr{\left[\sigma_0\tau_z\Sigma(0)
    \right]}}$, where $\Sigma^\mathrm{HF}$ is the limit of $\Sigma$ at high frequency: here the $\omega$-dependence is hardly visible and $\Sigma$ can be seen as that of a static Hartree-Fock-like mean field~\cite{Amaricci2015PRL,Amaricci2016PRB}. The
behavior of $\Xi$ is represented by the background color. 
In \figu{fig1}b the evolution of $\Xi$ as a function of $M$ is
smooth everywhere and, similar to the non-interacting case, the Weyl
nodes continuously meet in $k$-space at each insulator-to-semimetal
transition. 
In  \figu{fig1}c, the correlation strength changes discontinuously
at specific 
values of $M$ at both sides of the dashed line, signalling the abrupt
occurrence of the WSM phase from both directions. 
Since electron-electron interaction suppresses charge fluctuations associated with
the formation of  Weyl nodes, the WSM region generally shrinks as
$U$ increases. 
%The discontinuous evolution of $\Xi$ reflects the
%fact that the two monopoles associated with the Berry flux 
%appear suddenly and already separated in $k$, instead of evolving
%smoothly from a Dirac node. 

For even larger interaction strength  (Fig.~\ref{fig1}d), the system 
undergoes a discontinuous transition directly from the strong TI to the trivial BI 
without the semimetal in between. 
In this regime, the character of the WSM would be so strongly modified by many-body effects that the most natural evolution for the system is
to link the two distinct insulating phases without any intermediate
semimetal.  This is evident from the sudden change of color at
the critical value of $M$, which takes place with 
hardly any shading, as the semi-metallic region is completely
suppressed. To summarize, in the non-trivial regions the
groundstate develops a robust many-body character, encoded in a large (negative) value
of $\Xi$,  which cannot be continuously reconciled with the far
less correlated nature of the solution in the trivial BI phase. This
effect manifests in the discontinuous change of the (frequency
dependent) self-energy. This behavior is reminiscent of the strongly correlated transition found in the
presence of TRS and inversion symmetry, where a Dirac semimetal line (rather than a WSM region) cannot be continued
 to large values of $U$~\cite{Amaricci2015PRL,Amaricci2016PRB}.

There is a more intuitive way of interpreting our results. 
Concerning the topological nature of the system, 
the solution of the interacting problem can be recast into a
quadratic effective Hamiltonian containing all the terms of
Eq.~\ref{Hmodel}.  This way the effect of the electron-electron
interaction is accounted for by renormalized parameters whose
evolution can be compared to the non-interacting case. 
The two relevant indicators are the effective mass
$M^\text{eff}\! =\! M+\text{Tr}[\sigma_0\tau_z\Sigma(0)]/4$
and the effective TRS-breaking field
$b^\text{eff}_z \! =\! b_z + \text{Tr}[\sigma_z\tau_z\Sigma(0)]/4$. 
$M^\text{eff}$ controls the \emph{renormalized} energy separation between the 
orbitals, while $b^\text{eff}_z$ corresponds to the effective lifting
of the spin degeneracy. 
The behavior of these two quantities is displayed in \figu{fig2} as a
function of the bare mass term $M\!-\!M_0$, where $M_0$ is the critical value
of $M$ for which $M^\text{eff}\!=\!3$, i.e. the condition to have a
topological phase-transition for $b_z=0$~\cite{Amaricci2015PRL}.
% point. 
% the effective model described by the quadratic effective Hamiltonian displays a topological transition for $b_z\!=\!0$, analogously to its noniteracting counterpart \eqref{Hmodel}. 
In addition, we show the behavior
of the orbital compressibility $\kappa=\partial{T^z_i}/\partial{M}$,
where $T^z_i = \sum_\sigma n_{i1\sigma}-n_{i2\sigma} $ is the local
orbital polarization. 
In analogy with the non-interacting model, the WSM region is
characterized by $| M^\text{eff} -3 | \leq b^\text{eff}_z$ and is marked
in yellow. 

In the weak coupling regime (\figu{fig2}a) the effective mass evolves continuously
across the boundary of the WSM phase. The orbital compressibility
smoothly decreases upon approaching the BI region. 
\begin{figure}%[t]
  \begin{center}
    \includegraphics[width=1\linewidth]{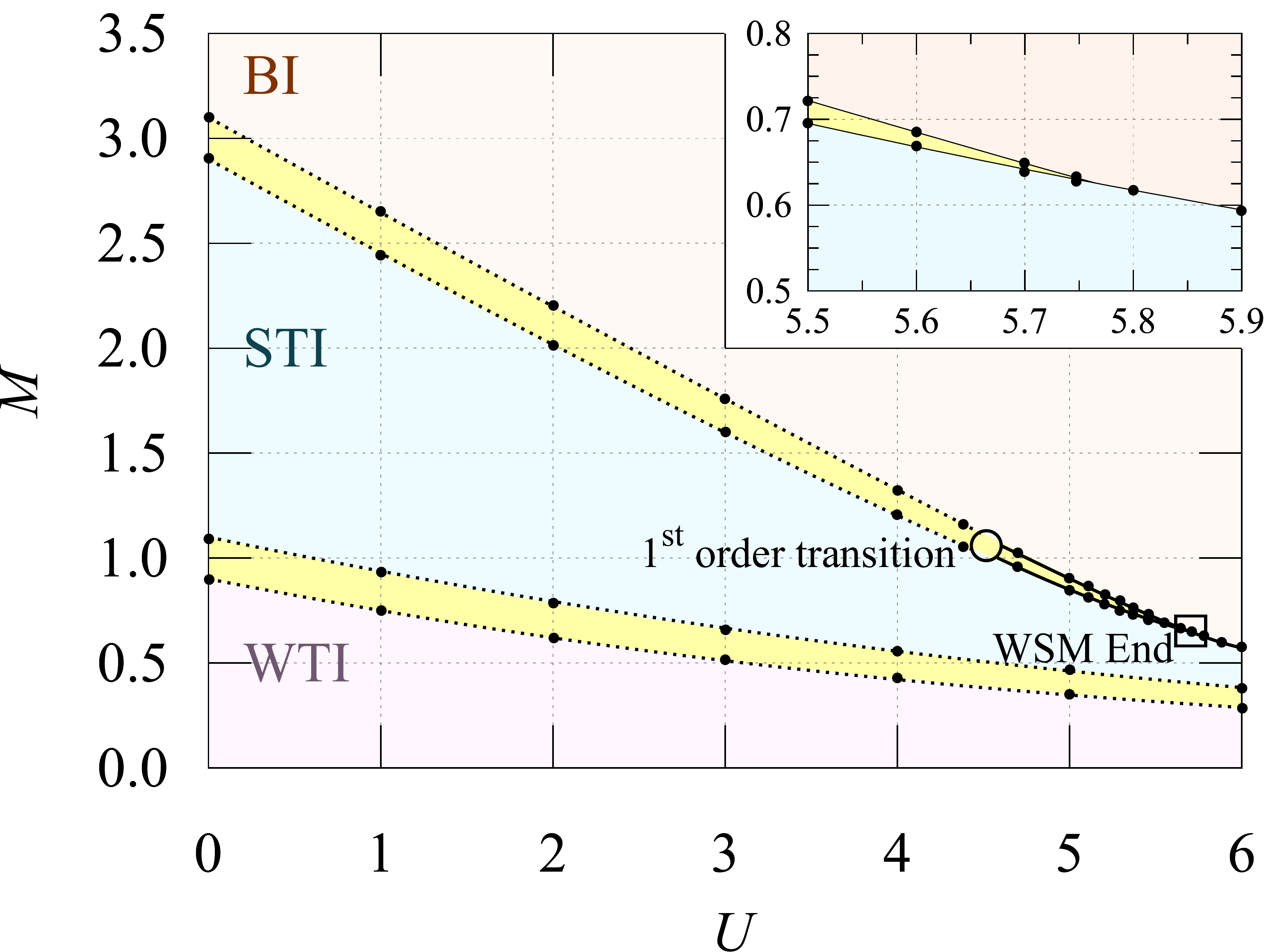}
    \caption{(Color online) 
    Phase diagram of the model for $b_{z}=0.1$. The regions (yellow) between the dotted lines are WSM. The inset shows the closing of the upper WSM region. Numerical results are represented by black symbols. At $U=0$, the
    widths of the semimetal regions are $2b_{z}$, centred around
    $M^\text{eff}=1,3$. The first semimetal region remains open
    when increasing the interaction, and the transition from WTI to
    WSM to STI is always continuous. A second critical point
    signals the value of $U$ for which the upper WSM region
    closes.}
%\lc{The behaviour of the effective parameters through the boundary lines, for the
%    values signalled by the dots, is of the type of \figu{fig2}, with the topological 	    phases denoted by the presence of nonzero chirality Weyl points for the semimetal and by 	    the usual 4-dimensional topological invariant \citep{Wang2012PRB} for the insulators}.
%    }
    \label{fig3}
\end{center}
\end{figure}
On the contrary, \figu{fig2}b the behavior of the effective
parameters becomes {\textit{discontinuous}}. 
In particular, the evolution of the effective mass term $M^\text{eff}$ exhibits
pronounced discontinuities at the crossing of the $b^\text{eff}_z$
lines delimiting the semi-metallic phase.  
The many-body nature of the strongly correlated solutions is 
demonstrated by the behavior of $\kappa$, displaying large peaks at
the transition points. Interestingly, the WSM region extends asymmetrically around  $M_0$. 
Indeed, the effect of the interaction is more pronounced in the strong TI
phase, because of the more fluctuating orbital polarization $T_z$, 
as compared to the close to fully orbitally polarized BI.   
In \figu{fig2}c, we report the evolution of the
effective parameters for a larger interaction strength $U$.  
In this regime, the Weyl region is entirely absent. $M^\text{eff}$ has a large
discontinuity just around the critical value $M_0$, unveiling the
direct first-order transition between the STI and the BI phases. This
transition is characterized by a large peak of the orbital
compressibility at the transition point.   
In the inset to \figu{fig2}b we show how the jumps in $M^\text{eff}$ evolve as a function of the temperature $T$.
The temperature at which they disappear for $U\!=\!5.5$ is of the order of $1/20$ of our hopping energy.

An overview of our findings is provided in \figu{fig3}, where we show the complete
phase diagram of the model in the $U$-$M$ plane. 
The different phases of the model are separated  by the two WSM
regions.
The boundary lines have an overall decreasing 
behavior as a  function of $U$, which results from the
tendency of the interaction to favour the orbitally polarized states,
i.e. the trivial BI with two electrons occupying the lowest orbital~\cite{Note1}. 
As anticipated above, the two WSM phases undergo a
dramatically different evolution in the presence of interaction. 
The WSM region  between WTI and STI is only slightly affected by
the electron-electron repulsion. This region displays a minor reduction of its
width and the transition lines keep their continuous character.   
On the other hand, the WSM region separating the STI and the BI
keeps its continuous character only up to a critical value of the
interaction $U_c=4.5$. Beyond this value the topological transition
lines acquire a first-order character. The 
width of this WSM phase gradually decreases upon increasing the 
interaction as a consequence of  the renormalization of the effective
term $b^\text{eff}_z$. 
In the inset of \figu{fig3} we 
show the evolution of the Weyl phase boundary lines up to the closure
of the corresponding interval for $U=5.75$.

Finally we investigate the fate of the boundary modes in the presence of correlations.
A distinctive property of  WSM are Fermi arcs connecting the projection of the two bulk Weyl nodes onto the 
surface Brillouin zone. In a single-particle scheme, approaching the 
topological transition is associated to the progressive shortening of the Fermi arcs
until the two ends coalesce in an insulator. It is natural to expect that the scenario 
we unveiled influences this picture.

\begin{figure}%[t]
	\begin{center}
	\includegraphics[width=1\linewidth]{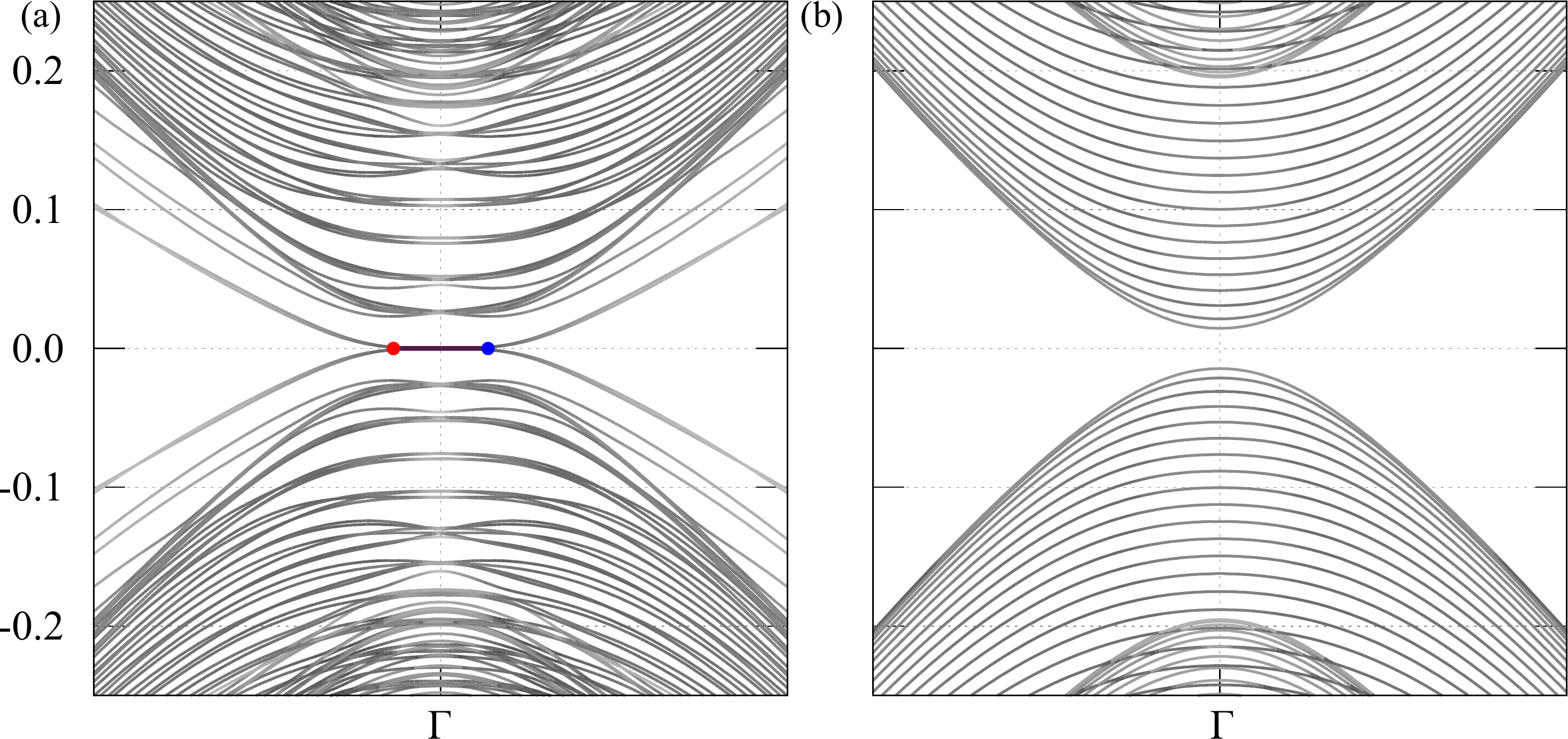}
	\caption{(Color online) Electronic structure of a slab
          geometry near the $\Gamma$ point at $U=5.6$. The
          two panels report the behavior for two different values of
          the mass term $M$ across the WSM to BI transition (see inset
          of \figu{fig3}). (a) $M=0.665$, the points (red and blue)
          indicate the Weyl nodes with opposite chirality. The band
          segment connecting them is the Fermi arc. (b) $M=0.670$, the system
          is a band insulator.}
	\label{fig4}
	\end{center}
\end{figure}

To address the correlated boundary states, in \figu{fig4} we show the spectrum of the model \equ{Hmodel} in 
a slab geometry with open boundary conditions along $y$. 
Our results illustrate the sudden (dis)appearance of the Fermi arcs across the
transition for a large value of the interaction $U$, reflecting the
first-order character of the transition observed in the bulk system
(see \figu{fig3}).

In conclusion, we revealed that in presence of a large electron-electron
interaction the transition to and from a WSM can become of first
order. In this scenario, the strongly
correlated Weyl nodes can  appear or annihilate 
discontinuously in a non-local fashion in reciprocal space at the
transition point. 
The change in the nature of the transition is inherently linked to a many-body
nature of the correlated Weyl semimetal whose distinctive property is 
an  enhancement of the orbital compressibility.  
We described this behavior by the evolution of effective model 
parameters. Finally, we showed that in the strongly interacting regime the
WSM region can progressively close in favor of a direct transition
between two gapped phases.

{\textit{Acknowledgements. }}
% \section{Acknowledgments}
The authors thank B.~Trauzettel for important comments and
careful reading of the manuscript.
A.A. and M.C. acknowledge support from H2020 Framework Programme, 
under ERC Advanced Grant No. 692670 ``FIRSTORM''. L.C., A.A. and M.C. acknowledge
financial support from  MIUR PRIN 2015 (Prot. 2015C5SEJJ001) and SISSA/CNR project 
``Superconductivity, Ferroelectricity and Magnetism in bad metals''
(Prot. 232/2015).
J.C.B. acknowledges financial support from the German Research
Foundation (DFG) through the Collaborative Research Centre SFB
1143. G.S. acknowledges financial support by the DFG through SFB 1170
``ToCoTronics'' and the W\"urzburg-Dresden Cluster of Excellence on
Complexity and Topology in Quantum Matter -- {\textit{ct.qmat}} (EXC
2147, project-id 39085490). N.~W. and G.~S. gratefully acknowledge the
Gauss Centre for Supercomputing
e.V. (\url{http://www.gauss-centre.eu}) for funding this project by
providing computing time on the GCS Supercomputer SuperMUC at Leibniz
Supercomputing Centre (\url{http://www.lrz.de}).

\bibliography{references}

\end{document}